\documentclass[conference]{IEEEtran}

\usepackage{tikz,tkz-euclide,pgfplots}
\tikzstyle{roundnode} = [circle, fill=black!255, scale=.8]
\newcommand{\roundlabel}[1]{\begin{tikzpicture}[baseline={([yshift=-0.8ex]current bounding box.center)}]\node[roundnode]{\color{white}\textbf{#1}};\end{tikzpicture}}
\tikzstyle{squarenode} = [square, fill=black!255, scale=1.0]

\usepackage{pifont}
\usepackage{amsmath}
\usepackage{balance}
\usepackage[noadjust]{cite}
\usepackage{threeparttable}
\usepackage{xcolor}
\usepackage{soul}           
\usepackage[linesnumbered, ruled, vlined]{algorithm2e}
\let\oldnl\nl
\newcommand{\nonl}{\renewcommand{\nl}{\let\nl\oldnl}}
\usepackage{comment}

\ifCLASSINFOpdf
  \usepackage[]{graphicx}
\else
\fi
\usepackage{array}
\usepackage{algcompatible}

\usepackage{multirow}
\usepackage{graphicx}
\usepackage{color, soul}
\usepackage{color, colortbl}

\usepackage{makecell}

\usepackage{breqn}

\begin{document}
\bstctlcite{IEEEexample:BSTcontrol}
%
\title{A Python Framework for SPICE Circuit Simulation of In-Memory Analog Computing Circuits}

\author{\IEEEauthorblockN{Md Hasibul Amin, Mohammed Elbtity, Ramtin Zand}
\IEEEauthorblockA{Department of Computer Science and Engineering, University of South Carolina, Columbia, SC 29208, USA
}
}



%


\maketitle

\IEEEpeerreviewmaketitle

\section{Introduction}

With the increased attention to memristive-based in-memory analog computing (IMAC) architectures \cite{9516756} as an alternative for energy-hungry computer systems for data-intensive applications, a tool that enables exploring their device- and circuit-level design space can significantly boost the research and development in this area. Thus, in this paper, we develop IMAC-Sim, a circuit-level simulator for the design space exploration and multi-objective optimization of IMAC architectures. IMAC-Sim is a Python-based simulation framework, which creates the SPICE netlist of the IMAC circuit based on various device- and circuit-level hyperparameters selected by the user, and automatically evaluates the accuracy, power consumption and latency of the developed circuit using a user-specified dataset. IMAC-Sim simulates the interconnect parasitic resistance and capacitance in the IMAC architectures, and is also equipped with horizontal and vertical partitioning techniques to surmount these reliability challenges \cite{parasitic_iscas}. In this abstract, we perform controlled experiments to exhibit some of the important capabilities of the IMAC-Sim.

\section{Proposed IMAC-Sim Framework}
\begin{figure}[t]
\centering
\includegraphics[width=3.4in]{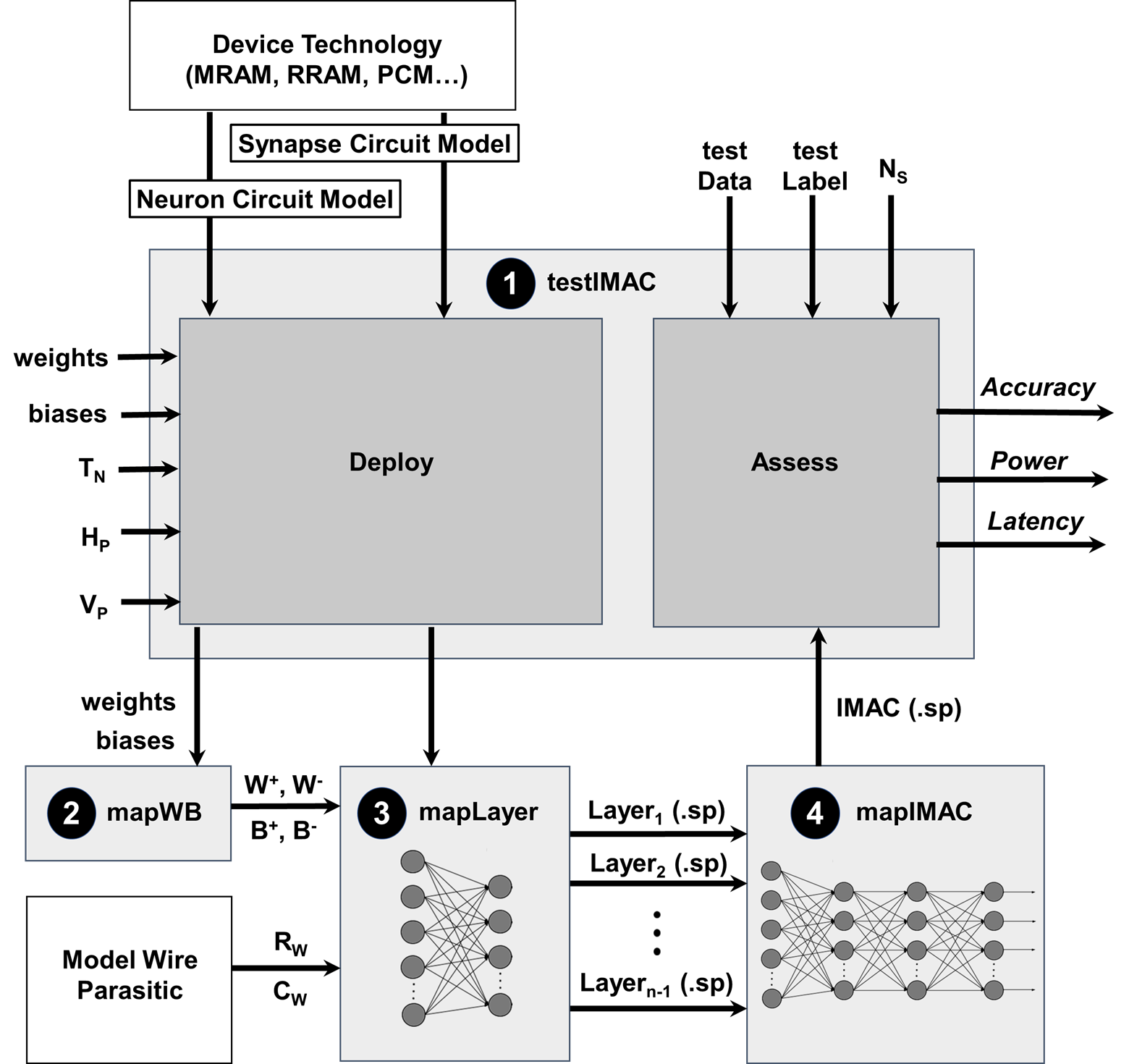}
\caption{Block diagram of the proposed IMAC-Sim framework.}
\label{fig:imacflow}
\end{figure}

Figure \ref{fig:imacflow} illustrates the structure of the IMAC-Sim framework, which includes four Python modules: \roundlabel{1} \textit{testIMAC}, \roundlabel{2} \textit{mapWB}, \roundlabel{3} \textit{mapLayer}, and \roundlabel{4} \textit{mapIMAC}. The \textit{testIMAC} module runs as a parent file, which controls the deployment of DNN workloads on IMAC architectures, as well as assessing their performance and accuracy. The \textit{testIMAC} module receives two sets of inputs from the user, as shown in Fig. \ref{fig:imacflow}. First, it takes trained $weights$ and $biases$, Network Topology ($T_N$), Horizontal Partitioning ($H_P$) and Vertical Partitioning ($V_P$) information, and the device- and circuit-level hyperparameters to deploy the DNN on the IMAC architecture. Next, it receives test dataset ($TestData$), test label ($TestLabel$), and Number of test samples ($N_S$) to assess the developed IMAC circuit. Table \ref{tab:hyperparameters} lists the hyperparameters of the IMAC-Sim.

\begin{table}[]
\caption{IMAC-Sim hyperparameters.}
\vspace{-2mm}
\centering
\begin{tabular}{lc}
\hline
Parameter               & Value    \\ \hline

Transistor Technology Node             & FinFET, CMOS     \\
Nominal Voltages         & [\textit{VDD}, \textit{VSS}]     \\
Neuron Circuit Model            & \textit{sigmoid}, \textit{tanh}, ReLU, \textit{etc.}          \\
Synaptic Technology        & [$R_{low}$ , $R_{high}$] 
\\
Network Topology   & $T_N$ = [$layer_1$, ..., $layer_{n}$]
\\
Vertical Partitioning   & $V_P$ = [$vp_1$, $vp_2$,..., $vp_{n-1}$]  \\
Horizontal Partitioning & $H_P$ = [$hp_1$, $hp_2$,..., $hp_{n-1}$]   \\
Differential Amplifier Gains             & [$G_1$, $G_2$,..., $G_{n-1}$] \\
Synapse Bitcell Size                     &     [\textit{Width}, \textit{Height}]             \\
Interconnect           & [\textit{resistivity}, \textit{thickness}, \textit{width}, ...]        \\
Sampling Time & $t_{sampling}$
\\

\hline
\end{tabular}
\label{tab:hyperparameters}
\end{table}

The functionality of \textit{testIMAC} module is demonstrated in Algorithm \ref{algo:testIMAC}. For each input sample, \textit{testIMAC} first stores the target labels in an array called $label$. It then calls another python module, \textit{mapIMAC}, which is responsible for creating the SPICE netlist of the IMAC circuit. For this purpose, \textit{mapIMAC} calls another python module, \textit{mapLayer}, which builds separate subcircuits for each of the layers in DNN including their interconnect parasitics and required partitioning as requested by user through $H_P$ and $V_P$ arrays. The \textit{mapLayer} modules returns the SPICE files for all of the layer subcircuits to \textit{mapIMAC}, which concatenates them to form the main \textit{IMAC} SPICE file. Finally, \textit{testIMAC} runs the SPICE simulation for the developed \textit{IMAC} SPICE file using the input voltages generated from the test dataset, and extracts the outputs of the last layer in IMAC circuit ($out$) and compares them with the $label$ to obtain the accuracy. Moreover, \textit{testIMAC} measures the average power consumption and latency of the circuit across various inputs and reports them to the user.

\begin{algorithm}
\small
\DontPrintSemicolon
\SetAlgoLined
\KwIn{test dataset ($TestData$), test label ($TestLabel$), $weights$, $biases$, Network Topology ($T_N$), No. of test samples ($N_S$), Horizontal Partitioning ($H_P$), Vertical Partitioning ($V_P$), $R_{low}$, $R_{high}$}
\BlankLine

 \textbf{Initialize:} $Error=0$, $PWR=0$, $T_N=[L_1, L_2, ...,L_n]$, $H_P=[hp_1, hp_2, ...,hp_{n-1}]$, $V_P=[vp_1, vp_2, ...,vp_{n-1}]$\;
 
 \nonl $/*\ Module\ \roundlabel{2}$\;
 W\textsuperscript+, W\textsuperscript-, B\textsuperscript+, B\textsuperscript- $\Leftarrow$ \textbf{mapWB} (\textit{weights}, \textit{biases})\;
 \For{$i=1$ \KwTo $N_S$}{
  \textit{label} $\Leftarrow$ target labels for input $i$\;
  \nonl $/*\ Module\ \roundlabel{4}$\;
  \SetKwProg{Fn}{}{:}{\KwRet {\textit{IMAC} SPICE circuit}}
  \Fn{\textbf{mapIMAC}{ ($T_N$, $H_P$, $V_P$, W\textsuperscript+, W\textsuperscript-, B\textsuperscript+, B\textsuperscript-)}}{

     \For{$j=1$ \KwTo $j=len(T_N)-1$}{
     \nonl $/*\ Module\ \roundlabel{3}$\;
     \SetKwProg{Fn}{}{:}{\KwRet {$Layer_j$ SPICE subcircuit}}
  \Fn{\textbf{mapLayer}{ ($hp_j$, $vp_j$, W\textsuperscript+, W\textsuperscript-, B\textsuperscript+, B\textsuperscript-)}}{
  Insert interconnection parasitics\;
  Implement the partitioning ($hp_j$, $vp_j$)\;
  }}
  Concatenate the $layer_j$ subcircuits\;
  }
  Run the SPICE simulation for \textit{IMAC} circuit\;
  \textit{out} $\Leftarrow$ Output of the IMAC circuit \;
  $PWR \mathrel{{+}{=}}$ power consumption of IMAC circuit\;
  
  \If{(out$\neq$label)}{
  $Error  \mathrel{{+}{=}} 1$;
  }
}
print $ErrorRate=Error/N_S$\;
print $P_{average}=PWR/N_S$\;
 \caption{\textit{IMAC-Sim} Framework}
 \label{algo:testIMAC}
\end{algorithm}

\section{Simulation Results and Discussion}

We utilize IMAC-Sim framework to implement a $400\times120\times84\times10$ DNN for classification application using an MNIST dataset. To limit the wide IMAC design space, we have fixed some of the hyperparameters as listed in Table \ref{tab:hyperpardefaults}.

\begin{table}[]
\caption{Hyper-parameters fixed herein to limit the IMAC design space.}
\vspace{-2mm}
\label{tab:hyperpardefaults}
\centering
\begin{tabular}{lc}
\hline
Parameter               & Value    \\ \hline

Transistor Technology Node             & 14 nm FinFET    \\ \hline
Nominal Voltages         & [\textit{VDD} $=0.8$V, \textit{VSS}$=-0.8$V]     \\ \hline
Neuron Circuit Model            & Memristive Sigmoid \cite{GLSVLSI_neuron}        \\ \hline
\multirow{2}{*}{Synapse Bitcell Size} & \textit{Width $=64\lambda=576 nm$} \\
                                & \textit{Height$=64\lambda=576nm$} 

\\ \hline

\multirow{3}{*}{Interconnect} &  \textit{Resistivity $\rho=1.9\times 10^9 \Omega.m$}  \\
                   & \textit{Thickness$=22  nm$} \\
                   & \textit{Width$=4\lambda= 36 nm$} \\ \hline

Inter-metal layer spacing & $20 nm$
\\

\hline
\end{tabular}
\end{table}



\subsection{Effect of Partitioning}

While IMAC-Sim supports any arbitrary value for horizontal and vertical partitioning, here we select the number of partitions per layer based on the maximum utilization of IMAC subarrays with various dimensions, as listed in Table \ref{tab:mram_part}. For instance, every layer in the  $400\times120\times84\times10$ DNN can be deployed on an IMAC architecture with $512\times512$ subarrays without partitioning, while the first layer requires to be divided into two horizontal partitions if we use $256\times256$ subarrays. 
The results listed in Table \ref{tab:mram_part} show that as the number of horizontal and vertical partitions increases both accuracy and power dissipation. In another test, we increased the number of partitions to $H_P=[16,8,8]$ and $V_P=[8,8,1]$, as listed in the last row of Table \ref{tab:mram_part}. Based on the results obtained from IMAC-Sim, this deployment scenario results in a high accuracy of 94.04\% at the cost of 60\% higher power dissipation. These types of trade-offs are important information that can be provided to developers by IMAC-Sim framework.

\begin{table}[]
\caption{IMAC array Partitioning results.}
\vspace{-2mm}
\label{tab:mram_part}
\centering
\begin{tabular}{ccccccccc}
\hline
\multirow{3}{*}{\begin{tabular}[c]{@{}c@{}}Array \\ Size\end{tabular}} & \multicolumn{6}{c}{Partitioning}                                                        & \multirow{3}{*}{Accuracy} & \multirow{3}{*}{\begin{tabular}[c]{@{}c@{}}Power\\  (W)\end{tabular}} \\ \cline{2-7}
                                                                       & \multicolumn{3}{c}{Horizontal ($H_P$)} & \multicolumn{3}{c}{Vertical ($V_P$)} &                           &                                                                                       \\ \cline{2-7}
                                                                       & L1          & L2         & L3         & L1         & L2         & L3        &                           &                                                                                       \\ \hline
                                                                      32$\times$32 & 13           & 4            & 3             & 4           & 3            & 1            & 73.64\%                   & 1.747                                                                               \\
                                                                      64$\times$64 & 7            & 2            & 2             & 2           & 2            & 1            & 28.44\%                   & 0.926                                                                              \\
                                                                      128$\times$128 & 4            & 1            & 1             & 1           & 1            & 1            &      11.35\%                      & 0.476                                                                              \\
                                                                      256$\times$256 & 2            & 1            & 1             & 1           & 1            & 1            & 11.35\%                          & 0.478                                                                              \\
                                                                        512$\times$512 & 1            & 1            & 1             & 1           & 1            & 1            & 11.35\%                          & 0.479                                                                              \\ \hline 
\rowcolor[HTML]{EFEFEF}
                                                                      32$\times$32 & 16           & 8            & 8             & 8           & 8            & 1            & \textbf{94.04}\%                   & \textbf{2.774}                                                                                 \\ \hline
\end{tabular}
\end{table}

\subsection{Effect of Memristive Device Technology}
We investigate the impact of memristive device technology on the performance of the IMAC architecture using four resistive technologies MRAM \cite{zand2018fundamentals}, RRAM \cite{li2018analogue}, CBRAM \cite{shi2018neuroinspired}, and PCM \cite{PCM-ratio}. Here, $R_{on}$ and $R_{off}$ values for different devices are changed for each run, while $H_P$ and $V_P$ are fixed to [13,4,3] and [4,3,1] respectively. Results listed in Table \ref{tab:tech_param} show that PCM-based IMAC architecture can achieve a high accuracy of 96.66\%, while consuming significantly less power compared to other technologies. This can be justified by the larger resistance of the PCM devices. 



\begin{table}[]
\caption{Impact of various memristive technologies on the accuracy and power consumption of IMAC architectures. }
\vspace{-2mm}
\label{tab:tech_param}
\centering
\begin{tabular}{ccccc}
\hline
Technology & $R_{low}$ & $R_{high}$  & Accuracy & Power ($W$) \\ \hline
MRAM \cite{zand2018fundamentals}      & $8.5 K\Omega$   & $25.5 K\Omega$        & 73.64\%  & 1.747           \\
 RRAM \cite{li2018analogue}       & $2.5 K\Omega$ & $100 K\Omega$       & 35.61\%  & 2.775          \\
 CBRAM \cite{shi2018neuroinspired}     & $5 K\Omega$   & $1 M\Omega$      & 69.56\% & 1.967            \\
  PCM \cite{PCM-ratio}      & $50 K\Omega$  & $1M\Omega$       & \textbf{96.66}\% & \textbf{0.447}          \\ \hline
\end{tabular}
\end{table}

\bibliographystyle{IEEEtran}

\balance
\bibliography{ref}
%



\end{document}